\documentclass[superscriptaddress,showpacs,aps]{revtex4}  

\usepackage{amsmath}
\usepackage{amssymb}
\usepackage{graphicx}
\usepackage{hyperref}

\newcommand{\dt}{\partial_t}
\newcommand{\vb}{{\bf v}}

\newcommand{\be}{\begin{equation}}
\newcommand{\ee}{\end{equation}}
\newcommand{\bean}{\begin{eqnarray}}
\newcommand{\eean}{\end{eqnarray}}

\begin{document}

\title{Chaotic dynamics in two-dimensional Rayleigh-B\'enard convection}

\author{Supriyo Paul}
\affiliation{Department of Physics, Indian Institute of Technology, Kanpur~208 016, India}

\author{Mahendra K. Verma}  
\affiliation{Department of Physics, Indian Institute of Technology, Kanpur~208 016, India}

\author{Pankaj Wahi}
\affiliation{Department of Mechanical Engineering, Indian Institute of Technology, Kanpur~208 016, India}

\author{Sandeep K. Reddy}
\affiliation{Department of Mechanical Engineering, Indian Institute of Technology, Kanpur~208 016, India}

\author{Krishna Kumar}
\affiliation{Department of Physics and Meteorology, Indian Institute of Technology, Kharagpur~721 302, India}

\date{\today} 

\begin{abstract} 
We investigate the origin of various convective patterns using bifurcation diagrams that are constructed using direct numerical simulations.   We perform two-dimensional pseudospectral simulations for a Prandtl number 6.8 fluid that is confined in a box with aspect ratio $\Gamma = 2\sqrt{2}$.   Steady convective rolls are born from the conduction state through a pitchfork bifurcation at $r=1$, where $r$ is the reduced Rayleigh number.  These fixed points bifurcate successively to time-periodic and quasiperiodic rolls through Hopf and Neimark-Sacker bifurcations at $r \simeq 80$ and $r \simeq 500 $ respectively.  The system becomes chaotic at $r \simeq 750$ through a quasiperiodic route to chaos. The size of the chaotic attractor increases at $r \simeq 840$ through an ``attractor-merging crisis'' which also results in travelling chaotic rolls.  We also observe coexistence of stable fixed points and a chaotic attractor for $ 846 \le r \le 849$ as a result of a subcritical Hopf bifurcation.  Subsequently the chaotic attractor disappears through a ``boundary crisis'' and only stable fixed points remain.  Later these fixed points  become periodic and chaotic through another set of bifurcations which ultimately leads to turbulence.
\end{abstract}

 

\maketitle

\section{Introduction}

Rayleigh-B\'enard convection (RBC), an idealized version of thermal convection, is studied for understanding pattern-forming instabilities, chaos, and turbulence in atmosphere, astrophysics, crystal growth, etc.~\cite{chandrashekhar}.   The flow dynamics in RBC is  governed by two nondimensional parameters: Rayleigh number  $R$, which is the ratio of  buoyancy and the dissipative terms, and the  Prandtl number $ P$, which is the ratio of the kinematic viscosity and the thermal diffusivity.  Convection starts at the critical Rayleigh number $R_c$ that has been found to be independent of the Prandtl number.  The numerical value of $R_c$ is approximately 1708 for no-slip boundary conditions, and 657.5 for free-slip boundary conditions~\cite{chandrashekhar}.    

Secondary instabilities and the resulting patterns of RBC depend quite critically on the Prandtl number.  For low Prandtl-number (low-P) and zero Prandtl number	(zero-P)	convection, the inertial term ${\bf u \cdot \nabla u}$ is quite important and it generates vertical vorticity.  Consequently the flow pattern becomes three-dimensional, and oscillatory waves along the horizontal axes are generated just near the onset of convection. For large Prandtl number (large-P) convection, however,  vertical vorticity is absent near the onset and  two-dimensional (2D) rolls survive till large Rayleigh numbers~\cite{chandrashekhar,schlutter:jfm_1965}.  This feature of large-P convection makes 2D simulations quite relevant for this regime.

Krishnamurti~\cite{krishnamurti:jfm_1970} carried out extensive convection experiments on fluids of different Prandtl numbers ($0.025 \le P \le 8500$) and studied various convective states such as two-dimensional, time-periodic, and chaotic rolls,  and turbulent structures.    Gollub and Benson~\cite{gollub:jfm_1980} performed RBC experiments on water for $P=2.5$ and 5.0 (varied by changing the mean temperature) and aspect ratios $\Gamma = 2.4$ and $3.5$.  They observed a variety of convective states and several routes to chaos.  For $\Gamma=3.5$ and $P=5.0$, they found a quasiperiodic state at the reduced Rayleigh number $r=R/R_{c} \simeq 32$ and subsequent phase locking at $r \simeq 44.4$.  The flow became chaotic at around $r \simeq 46$, thus exhibiting a quasiperiodic route to chaos.  Gollub and Benson~\cite{gollub:jfm_1980} also observed a period-doubling route to chaos for $\Gamma=3.5$ and $P=2.5$, and a quasiperiodic route to chaos with three frequencies (Ruelle-Takens) for $\Gamma=2.4$ and $P=5.0$. Intermittent chaos too was observed in their experiments.   Maurer and Libchaber~\cite{Maurer:1979} observed quasiperiodic and frequency-locking route to chaos in helium ($P \sim 0.7$). Experiments of Giglio {\it et al.}~\cite{Giglio:1981} on the convection of water ($P \sim 7$) showed a period doubling route to chaos. Berg\'e {\it et al.}~\cite{berge:jpl_1980} observed intermittency in their RBC experiments on silicon oil ($P \sim 100$).   Libchaber {\it et al.}~\cite{Libchaber} studied convection in mercury ($P \sim 0.02$) in the presence of an applied mean magnetic field and observed generation of chaos through period doubling and quasiperiodic routes.

Direct numerical simulations (DNS) are used extensively to study convection.  Curry et al.~\cite{curry:jfm_1984} performed DNS in three dimensions for a $P=10$ fluid under free-slip boundary conditions.  They reported steady convection till $r \simeq 40$, after which a limit cycle is observed till $r \simeq 45$.  Subsequently they reported quasiperiodicity ($r \simeq 45-55$), phase locking ($r \simeq 55-65$), and chaos ($r \ge 65$).  Yahata~\cite{Yahata:DNS} observed a similar route to chaos in his DNS for  $P=5$ , $\Gamma_x = 3.5$ and $\Gamma_y = 2$.  Mukutmoni and Yang~\cite{Mukutmoni:JHT_1993} performed DNS for convection in  a rectangular enclosure with insulated side walls and found a period-two state, but a quasiperiodic route to chaos.  Yahata~\cite{Yahata:DNS} and Nishikawa and Yahata~\cite{NishiYahata:DNS} performed direct numerical simulations of RBC for the setup similar to Gollub and Benson's~\cite{gollub:jfm_1980} experiments and observed quasiperiodic route to chaos. Yahata~\cite{Yahata:DNS}  also observed frequency locking.  Gelfgat~\cite{Gelfgat} numerically studied the effects of the aspect ratio on the critical Rayleigh number and the instability modes.

	Three-dimensional DNS are quite expensive computationally, so a large number of two-dimensional DNS have been performed. Moore and Weiss~\cite{moore_weiss:jfm_1973}  simulated for P = 6.8 using a finite-difference method under free-slip boundary conditions and reported that the Nusselt number $Nu \approx 2\, r^{1/3}$ for $5 \le r \le P^{1.5}$,  and $Nu \approx 2\, r^{0.365}$ for higher Rayleigh numbers.   McLaughlin and Orszag~\cite{mclaughlin_orszag:jfm_1982} simulated RBC in air (P = 0.71) with no-slip boundary conditions and observed periodic, quasiperiodic, and chaotic states for Rayleigh numbers between 6500 and 25000. Curry et al.~\cite{curry:jfm_1984} performed detailed DNS for $P = 6.8$  in 2D and observed oscillations with a single frequency at $r \simeq 50$, and with two frequencies (quasiperiodic) at $r \simeq 290$. They observed weak chaos beyond $r \simeq 290$, and a periodic convective state after $r\simeq 800$.  Goldhirsch et al.~\cite{goldhirsch:jfm_1989} also simulated 2D RBC and observed complex behaviour.  Zienicke~{\em et al.}~\cite{Zie}  studied the effects of symmetries on 2D RBC using DNS, and observed periodic, quasiperiodic, and phase-locked states.   

Simulations of convection reveals various convective patterns, yet the large number of modes present in DNS obscures the origin of these patterns.  Low-dimensional models, constructed using Galerkin projections, are very useful for investigating these issues.  Some of the earlier investigations using low-dimensional models are by Lorenz~\cite{Lorenz}, Mclaughlin and Martin~\cite{Mclaughlin:pra1975} and Curry~\cite{Curry:CommMath}. They observed various convective states including chaos through period doubling and quasiperiodic routes. Yahata~\cite{Yahata:lowD_3freq,Yahata:lowD_period_doubling} constructed a 48-mode model that is inspired by the experimental configurations of Gollub and Benson~\cite{gollub:jfm_1980} and reported quasiperiodic (Ruelle-Takens type) and period-doubling route to chaos.   Recently Paul {\it et al.}~\cite{Paul:highP} constructed a 30-mode model of 2D convection and studied it for $P=6.8$ and $\Gamma = 2 \sqrt{2}$; they observed transition from periodic $\rightarrow$ quasiperiodic $\rightarrow$  phase-locked state $\rightarrow$  chaos, similar to one of the  experimental results of Gollub and Benson~\cite{gollub:jfm_1980}.  

In this article we investigate the origin of various convective patterns using bifurcation diagrams constructed using DNS.  We adopt pseudospectral method to simulate convection in water at room temperature ($P=6.8$) in a box with aspect ratio of $\Gamma_x = 2\sqrt{2}$.  Bifurcation analysis requires a large number of computer runs.  Since 3D simulations are computationally very expensive, we resort to 2D simulations.  We obtain steady, periodic, quasiperiodic, and chaotic rolls as reported earlier by Curry {\it et al.}~\cite{curry:jfm_1984}, Yahata~\cite{Yahata:DNS}, and Mukutmoni and Yang~\cite{Mukutmoni:JHT_1993}.  In addition, we observe that the flow becomes steady after chaos.  Subsequently this steady state turns periodic and turbulent.  We also observe coexistence of fixed points and chaotic attractors.  We explain these features using bifurcation diagrams.

The outline of the paper is as follows.  In section II we describe the governing equations and the numerical method.  Section III contains descriptions of various convective states resulting from various bifurcations. In section IV we discuss the dynamics of the chaotic state. We describe the properties of the large scale modes in section V. The last section contains conclusions.   


\section{Hydrodynamic system and numerical method}

We consider a layer of Boussinesq fluid~\cite{chandrashekhar} of thickness $d$, kinematic viscosity $\nu$, thermal diffusivity $\kappa$, and thermal expansion coefficient $\alpha$ confined between two stress-free and thermally conducting horizontal plates. An adverse temperature  gradient $\beta=\Delta{T}/d$ is imposed across the fluid layer, where $\Delta{T}$ is the temperature difference between the plates. The relevant hydrodynamic equations are nondimensionalized by choosing 
length scale as $d$,  velocity scale  as $\sqrt{\alpha\beta g d^{2}}$, and temperature scale as $\Delta{T}$, which yields
\bean
\dt\vb + (\vb\cdot\nabla)\vb &=& -\nabla{p}+\theta\hat{z}+\sqrt{\frac{P}{R}}\nabla^2 \vb, \label{eq:NS}\\
\dt\theta + (\vb\cdot\nabla)\theta &=& v_3 + \frac{1}{\sqrt{PR}}\nabla^2 \theta, \label{eq:temp} \\
\nabla \cdot \vb & = & 0, \label{eq:continuity} 
\eean
where $\vb=(v_1,v_2,v_3)$  is the velocity fluctuation, $\theta$ is the perturbations in the  temperature field from the steady 
conduction state, $R=\alpha g \beta d^4/\nu \kappa$ is the Rayleigh number,  $P=\nu/\kappa$ is the Prandtl number, $g$ is the acceleration due to gravity, and $\hat{z}$ is the buoyancy direction. The two-dimensional rolls formed in the system are assumed to be parallel to the $y$ axis. The top and bottom boundaries are considered to be stress free and perfectly conducting: 
\bean
v_3 = \partial_{z}v_1 =\partial_{z}v_2 =  \theta = 0, ~~~~ \mbox{at}~~  z = 0, 1.  \label{eq:bc}
\eean
The velocity and temperature fields are periodic along the horizontal direction ($x$ axis).

The above set of equations (\ref{eq:NS}-\ref{eq:continuity})  are solved numerically using  a pseudospectral method~\cite{Canuto} in two dimensions under the above boundary conditions.  We use Fourier basis functions for representation  along the $x$ direction, and $\sin$ or $\cos$ functions for representation  along  the $z$ direction.  The velocity component along  the $y$-direction ($v_2$) is zero. The expansion of the velocity and temperature fields are
\bean
	v_1 (x, z, t) &=&  \sum_{m,n} 2 U_{m0n} (t) \exp(i m k_c x) \cos(n\pi z),\nonumber\\
	v_2 (x, z, t)  &=&  0,\nonumber\\
	v_3 (x, z, t)  &=&  \sum_{m,n} 2 W_{m0n} (t) \exp(i m k_c x) \sin(n\pi z), \nonumber\\
	\theta (x, z, t) &=&  \sum_{m,n}  2 \theta_{m0n} (t) \exp(i m k_c x) \sin(n\pi z),
	\label{eq:four_expansion}
\eean
where $k_c = \pi/\sqrt{2}$. 

Various grid resolutions, $64 \times 64$, $128\times 128$, $256\times 256$, $512\times 512$ have been used in
the simulations.  The aspect ratio of our simulations is  $\Gamma = 2\sqrt{2}$. Time stepping is carried out using the fourth-order 
Runge-Kutta (RK4) method with CFL scheme for choosing $dt$. The simulations are carried out till the system reaches a steady-state.  

We perform numerical simulations for  $1.01 < r < 5\times10^5$ ($R=664$ to $3.3\times 10^{8}$).  The thermal Prandtl number is chosen as $P = 6.8$ which is a typical value for water at room temperature.  In the next section we will describe various convective states obtained in our simulations and investigate their origin using bifurcation diagrams.


\section{Bifurcation diagrams for 2D RBC}
\label{time_series_study}

\begin{figure}[t]
\begin{center}
\includegraphics[height=!,width=16cm]{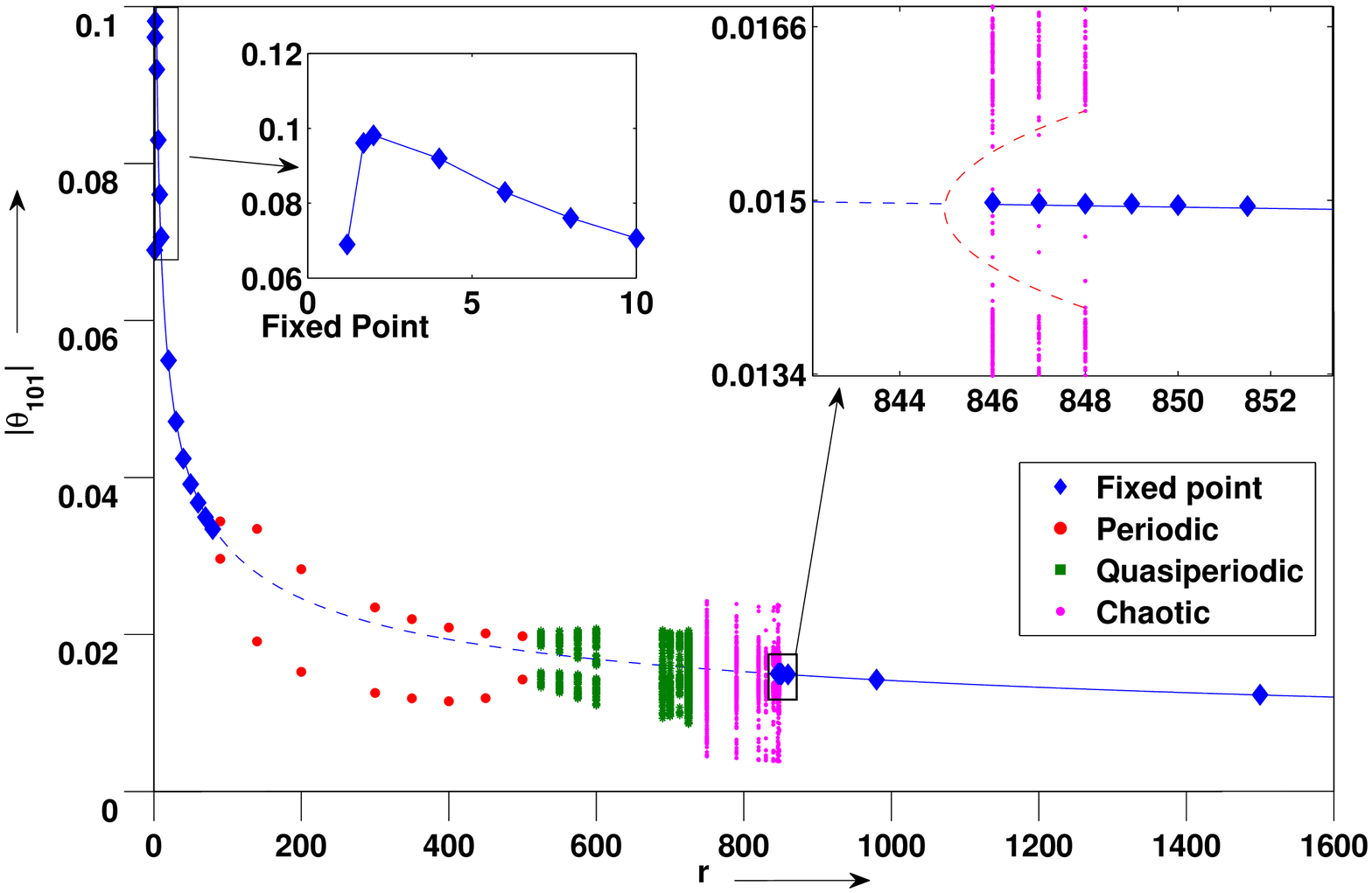}
\end{center}
\caption{Bifurcation diagram $|\theta_{101}|$ vs. $r$ computed using DNS results.  The system undergoes a pitchfork bifurcation at $r=1$. The new stable fixed points (blue diamonds) undergo a supercritical Hopf bifurcation at $r \simeq 80$ yielding stable time-periodic flows (red circles). Quasiperiodic solutions are born at $r \simeq 500$.  Subsequently chaos is observed in the band $750 \le r \le 849$.  Coexisting attractors, stable fixed points and a chaotic attractor, are observed for $846 < r< 849$ as shown in the right inset.   Subsequently the chaotic attractor disappears through a ``boundary crisis'' at $r=850$, and only stable fixed points survive.  }
\label{fig:bif_theta}
\end{figure}

\begin{figure}[t]
\begin{center}
\includegraphics[height=!,width=16cm]{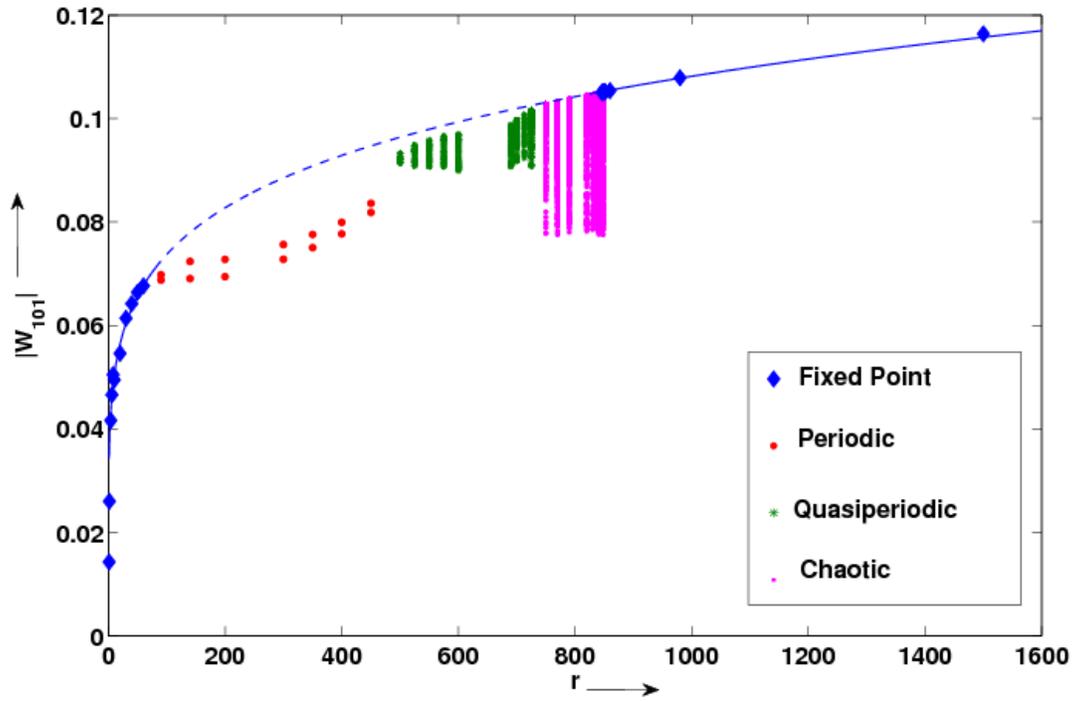}
\end{center}
\caption{Bifurcation diagram $|W_{101}|$ vs. $r$ computed using DNS results.  The symbols used for various states and their interpretations are same as Fig.~\ref{fig:bif_theta}.  }
\label{fig:bif_W}
\end{figure}

The number of variables of simulations on a $N \times N$ grid are $N^2$, which is quite large.  It is impossible to understand the system dynamics in terms of all these variables.  Fortunately numerical simulations reveal that some of the large-scale modes have large fraction of kinetic energy and entropy ($\int d{\bf x} |\theta|^2/2$) up to the chaotic regime.  Therefore we analyze various convective states of RBC using these large-scale modes, namely $\theta_{101}$ and $W_{101}$.  In Figs.~\ref{fig:bif_theta} and \ref{fig:bif_W} we plot the absolute values of these modes as a function of $r$.  For time-varying states like time-periodic and chaotic flows, we plot the minimum and maximum values of these variables.   The above figures, also called the bifurcation diagrams, have been created from the results of around sixty DNS runs.  In the following discussions we will describe the bifurcation diagrams in more detail.

\begin{figure}[t]
\begin{center}
\includegraphics[height=!,width=16cm]{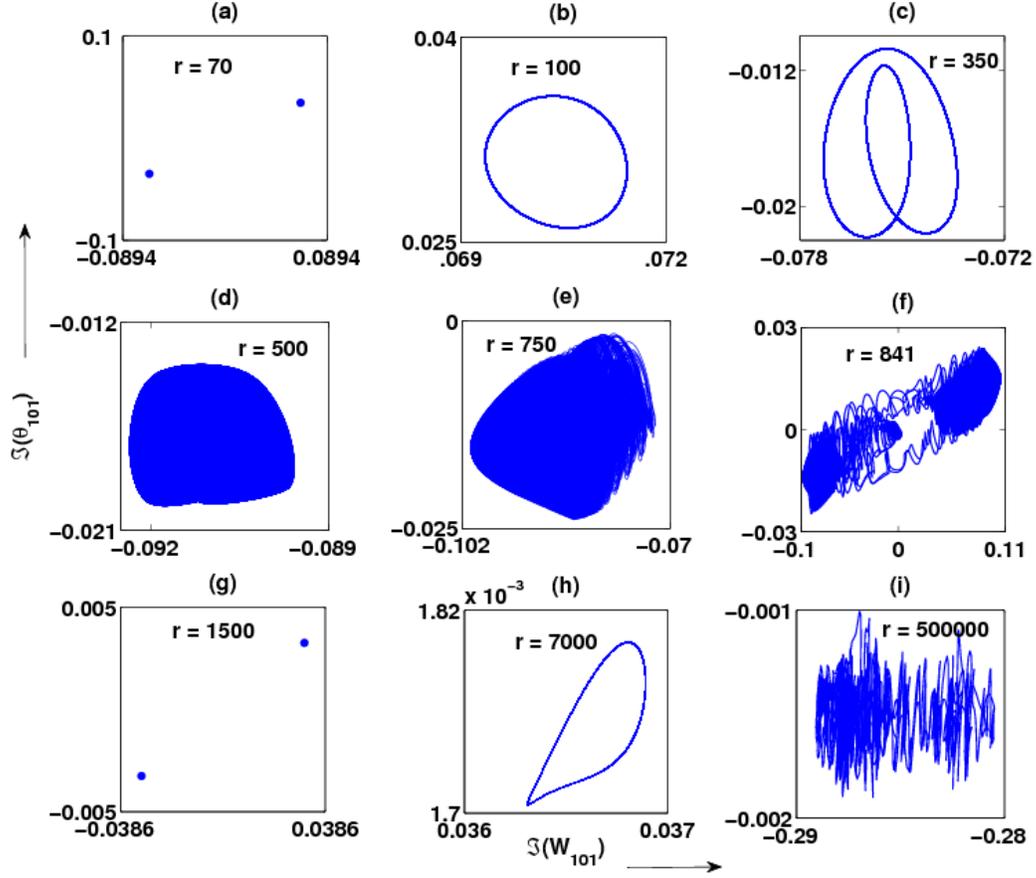}
\end{center}
\caption{For $P=6.8$, phase space projections on $\Im(W_{101})$ - $\Im(\theta_{101})$ plane of various convective states:  steady convective rolls for $r=70$ (a);  time-periodic state for $r=100$ (b);  time-periodic roll with period-2 behavior for $r=350$ (c); quasiperiodic state for $r=500$ (d);  chaotic states for $r=750$ (e) and for $r=841$ (f);  reemergence of steady convective rolls for $r=1500$ (g);   periodic state for $r=7000$ (h); turbulent state for $r=5\times10^{5}$ (i).}
\label{fig:phasespace}
\end{figure}

The phase space or state space of a system provides valuable information on the dynamics of that system.   The phase space of our dynamical system is $N^2$ dimensional, which is impossible to visualize.  In the following discussion we will show only a projection of the phase space onto a subspace made by some of the energetic modes.  In Fig.~\ref{fig:phasespace} we plot a phase space projection onto $\Im(W_{101})-\Im(\theta_{101})$ plane, where $\Im$ stands for the imaginary part of the argument.

The RBC system is in a conduction state for $r<1$. At $r=1$, the conductive state becomes unstable, and a convective state is born via a supercritical pitchfork bifurcation. A new stable solution, the stationary straight rolls, represented by blue diamonds in Figs.~\ref{fig:bif_theta} and \ref{fig:bif_W} emerge from the bifurcation point.  The stationary straight rolls exist till  $r \simeq 80$. A phase space projection of the fixed points  is shown in Fig.~\ref{fig:phasespace}(a).  Various secondary bifurcations take place after this primary bifurcation.  Before we proceed to discuss various convective states, we make an important remark regarding  the fixed point solutions and related states.  When the initial conditions of the velocity and temperature Fourier modes are chosen as imaginary (except only for $\theta_{00n}$ which is purely real), the fixed points are purely imaginary.  However, when we choose complex Fourier modes as initial conditions, we get fixed points with complex Fourier modes, but the absolute values of all the Fourier modes match with the magnitude of the above mentioned imaginary values.  This feature is a consequence of the translational invariance of the solutions in the $x$-direction  due to the periodicity of the box along $x$.  This manifests itself as a phase shift of the various Fourier modes.  Therefore a  purely imaginary mode can become complex with the same magnitude.  This symmetry plays an important role in the chaotic regime as well; these results will be described in our later discussions. 

\begin{figure}[t]
\begin{center}
\includegraphics[height=!,width=16cm]{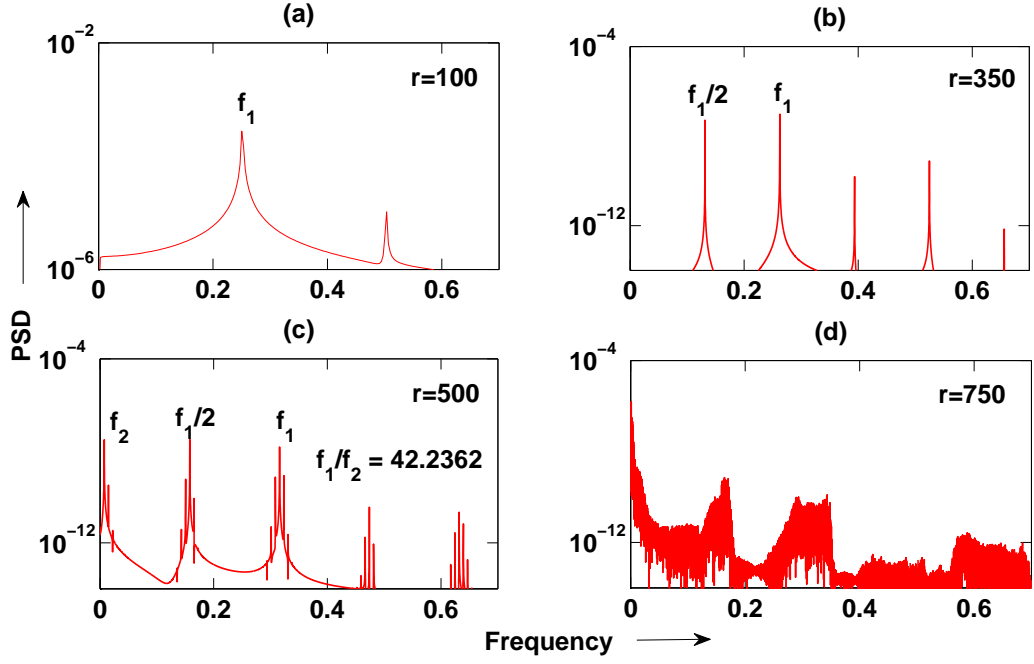}
\end{center}
\caption{Power spectrum of $\Im(W_{101})$ for various convective states: (a) for periodic state with dominant peak at  $f_1 = 0.2632$ ($r=100$); (b) for period-2 state with dominant peaks at $f_1/2$ and $f_1$ ($r=350$); (c) for quasiperiodic state with dominant peaks at $f_1/2$, $f_1$, and $f_2$ with $f_1/f_2 \approx 42.2362$ ($r=500$); (d) for chaotic state at $r=750$ with broadband spectrum. }
\label{fig:fft}
\end{figure}

Near $r\simeq 80$, the system bifurcates from a steady convection state to a periodic state (a limit cycle in the phase space) through a Hopf bifurcation.  The extremum values of $|W_{101}|$ and $|\theta_{101}|$ are shown in Figs.~\ref{fig:bif_theta} and \ref{fig:bif_W} respectively using red dots.  A sketch of the unstable fixed point solutions obtained using extrapolation is also shown in the bifurcation diagrams as a dashed blue curve.  Figure ~\ref{fig:phasespace}(b) exhibits a phase space projection of a limit cycle on the $\Im(W_{101})$ - $\Im(\theta_{101})$ plane for $r=100$.   The power spectrum of the time series of the Fourier mode  $\Im(W_{101})$ for the same $r$ is shown in Fig.~\ref{fig:fft}(a)  clearly depicting sharp peaks at $f=f_1$ (dominant) and $2 f_1$ (superharmonic) with $f_1 \approx 0.2624$. As $r$ is increased further, the periodic state undergoes a period-doubling bifurcation at around $r \simeq 350$, similar to that observed by Mukutmoni and Yang~\cite{Mukutmoni:JHT_1993}. As a consequence of this bifurcation, the phase space projection of the limit cycle on the $\Im(W_{101})$ - $\Im(\theta_{101})$ plane shows a second loop as depicted in Fig.~\ref{fig:phasespace}(c) for $r=350$.  For this state, the power spectrum acquires a peak at $f_1/2$ along with the dominant peak at $f_1$ as shown in Fig.~\ref{fig:fft}(b). Superharmonic frequencies are also present in Fig.~\ref{fig:fft}(b) due to the nonlinearity in the system. We note at this point that the period-2 behavior is not very apparent in the bifurcation diagrams (Figs.~\ref{fig:bif_theta} and \ref{fig:bif_W}) for the absolute values of the modes $W_{101}$ and $\theta_{101}$. This is possibly due to some cancellations while taking the absolute values of the modes.

The period-2 limit cycles described above become unstable at around $r\simeq 500$, and another independent frequency is born through a Neimark-Sacker bifurcation.  The power spectrum of $\Im(W_{101})$, shown in Fig.~\ref{fig:fft}(c) for $r=500$, indicates three dominant peaks at frequencies $f_1/2\simeq 0.158$, $f_1\simeq 0.316$ and $f_2 \simeq 0.0075$. The first two frequencies correspond to the period-2 limit cycle, while the third frequency is an irrational multiple of the first two ($f_1/f_2 \approx 42.2362$).   As a consequence,  the RBC state  for $r=500$ is quasiperiodic, and the phase space trajectories fill a torus in higher dimensions, whose projection on a plane is shown in Fig.~\ref{fig:phasespace}(d).  The quasiperiodic flows, observed for $500 \le r \le 725$,  are represented by green dots in the bifurcation diagrams (Figs.~\ref{fig:bif_theta}, \ref{fig:bif_W}).

At $r \simeq 750$, the quasiperiodic state eventually bifurcates to a chaotic state that  persists till $r \simeq 849$. In Figs.~\ref{fig:bif_theta} and \ref{fig:bif_W} the chaotic flows are illustrated as pink dots.   A phase space projection of a chaotic state at $r=750$ is shown in Fig.~\ref{fig:phasespace}(e).  The power spectrum for this state, shown in (Fig. \ref{fig:fft}(d)),  shows a broadband power spectra, a characteristic of a chaotic solution. This chaotic attractor (Fig.~\ref{fig:phasespace}(e)) observed in the range $750 \le r < 840$, is born through a quasiperiodic route to chaos.   This transition could either be through phase-locking or through Ruelle-Takens scenario~(\cite{berge:book,hilborn:book}), which can be ascertained only by performing simulations for many $r$ values in the neighbourhood.   DNS cannot help us ascertain which of the two scenario is applicable since simulations for many $r$ values in the neighbourhood are not possible.  Yet close similarity of DNS results with the corresponding low-dimensional calculations~\cite{Paul:highP} suggests that the transition is possibly through the phase-locking route. 

In the band $750 \le r \le 840$ we observe two unconnected chaotic attractors that lie in the first and the third quadrants of the phase space projection. That is, both $\Im(W_{101})$ and  $\Im(\theta_{101})$ either take positive values or negative values.  At later $r$ values, for $841 < r \le 849$, these two disjoint attractors merge through an ``attractor merging crisis'' ~\cite{hilborn:book, Grebogi:PRA_1987} to create a larger chaotic attractor shown in Fig. \ref{fig:phasespace}(f). However the increase in the size of the chaotic attractor is not apparent in the bifurcation diagrams (Figs.~\ref{fig:bif_theta} and \ref{fig:bif_W}) due to the use of absolute values of these modes for depiction.  

\begin{figure}[t]
\begin{center}
\includegraphics[height=!,width=16cm]{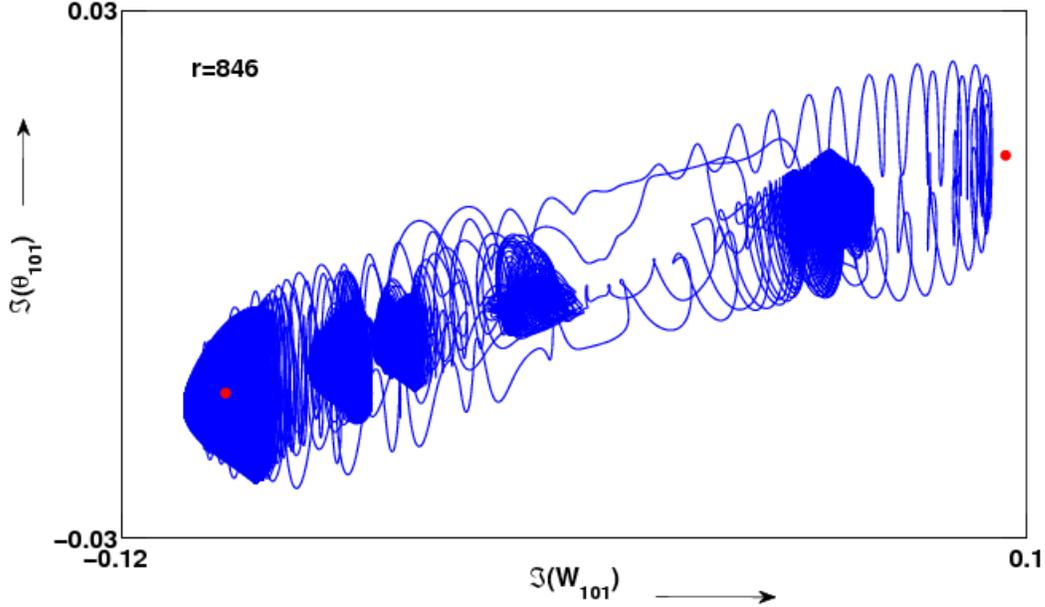}
\end{center}
\caption{ For $r=846$, a phase space projection of coexisting attractors, stable fixed points (red dots) and a chaotic attractor (blue trajectory).  }
\label{fig:coexist_attractor}
\end{figure}

As $r$ is increased further, we observe two coexisting attractors in the window of $846 \leq r \leq 849$.  For some initial conditions we observe chaos, and for some others we find stable fixed points. For $r=846$, the coexisting attractors in a phase space projection are shown in Fig.~\ref{fig:coexist_attractor} wherein the dots represent stable fixed points, while the trajectory corresponds to a chaotic solution. This feature can be understood as follows. The unstable fixed points (dashed blue line in the bifurcation diagram) undergoes an inverse subcritical Hopf bifurcation at $r \simeq 845$.  As a result, an unstable limit cycle is born (dashed red lines in the inset of Fig.~\ref{fig:bif_theta}), which coexists with the stable fixed points (the solid blue line in the inset) for $r>845$. As a result of this bifurcation, stable fixed points and a chaotic attractor are observed to coexist for  $846 \leq r \leq 849$ with the unstable limit cycle (along with its stable manifold) approximately forming the basin boundary between the two.   

\begin{figure}[t]
\begin{center}
\includegraphics[height=!,width=16cm]{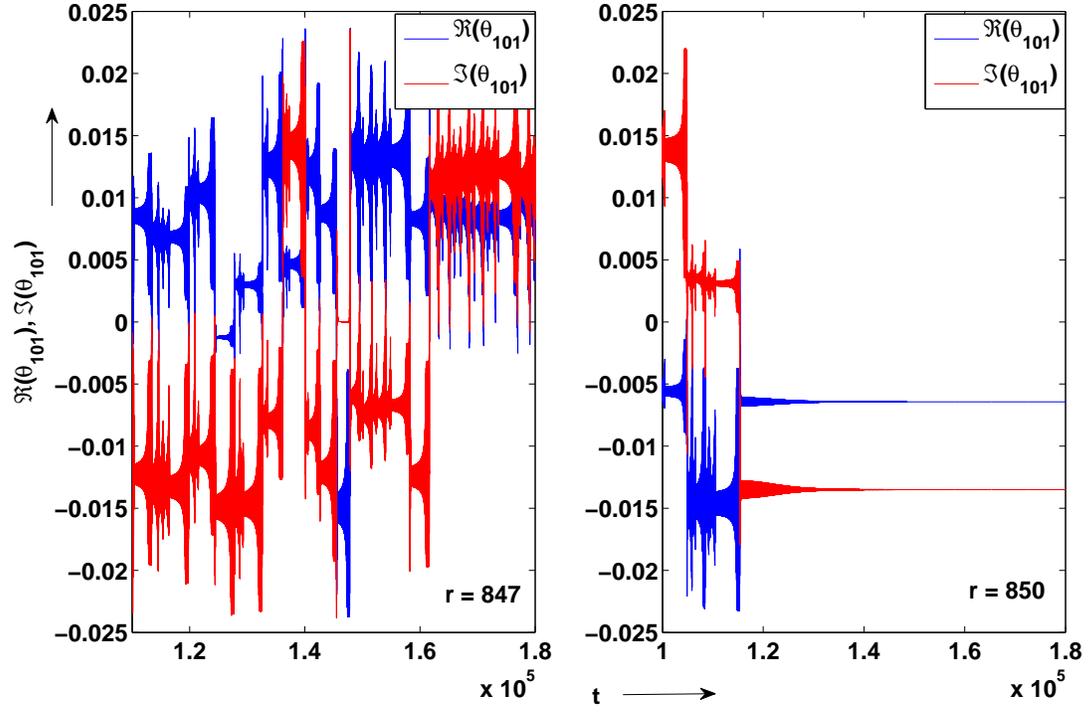}
\end{center}
\caption{Time series of $\Re(\theta_{101})$ and $\Im(\theta_{101})$ for $r=847$ and $r=850$. At $r=847$ the system is chaotic, while at $r=850$ the system settles down to a fixed point solution after chaotic transients.}
\label{fig:crisis}
\end{figure}

At $r\simeq 849$, the chaotic attractor disappears, and we observe only stable fixed points as depicted in Fig.~\ref{fig:phasespace}(g) for $r=1500$.  The disappearance of the chaotic attractor is  due to a ``boundary crisis''~\cite{hilborn:book, Grebogi:physica_1983} where the chaotic attractor collides with the unstable limit cycle.  Figure~\ref{fig:crisis} shows the time series of $\Re(\theta_{101})$ and $\Im(\theta_{101})$ for $r=847$ and $850$.  The system is chaotic for $r=847$, however for $r=850$ it settles down to  a stable fixed point after chaotic transients.   A close similarity between the long transients for $r=850$ and the chaotic time series for $r=847$ provides an evidence for the occurrence of a ``boundary crisis''. 

The new fixed points continue to remain stable for a significantly large range of $r$ values.  At higher $r$ values, the fixed points bifurcate to  limit cycles through a Hopf bifurcation.  A phase space projection of a limit cycle for $r=7000$ is shown in Fig.~\ref{fig:phasespace}(h).  At later $r$ values, the limit cycles bifurcate to a turbulent state, similar to that reported by Vincent and Yuen~\cite{Vincent}.  A sample  phase space projection for this range is shown in Fig.~\ref{fig:phasespace}(i) for $r=5\times 10^5$.   The properties of turbulent state has been explored earlier by Vincent and Yuen~\cite{Vincent} for $ 10^8 < r< 10^{14}$, and they are not discussed in this paper.  Our bifurcation diagrams shown in Figs. \ref{fig:bif_theta} and \ref{fig:bif_W} are quite detailed till $r=1600$.

The quasiperiodic route to chaos for 2D RBC presented here is in general agreement with those described earlier in some experiments~\cite{gollub:jfm_1980}, simulations~\cite{curry:jfm_1984,Yahata:DNS,Mukutmoni:JHT_1993,NishiYahata:DNS,mclaughlin_orszag:jfm_1982}, and low-dimensional models~\cite{Paul:highP}.  There is a similarity with the RBC simulations  of  Mukutmoni and Yang~\cite{Mukutmoni:JHT_1993}, who observed a period-two state but subsequent route to chaos through quasiperiodicity.    We report for the first time stable fixed points beyond chaos, and coexistent stable fixed points and chaotic states.   We also observe several ``crisis'' for the first time in 2D RBC simulations: first, smaller attractors corresponding to Fig.~\ref{fig:phasespace}(e) merge to form a larger attractor shown in Fig.~\ref{fig:phasespace}(f) through an ``attractor-merging crisis'', and second, the disappearance of the chaotic attractor through a ``boundary crisis''.   

\begin{figure}[t]
\begin{center}
\includegraphics[height=!,width=16cm]{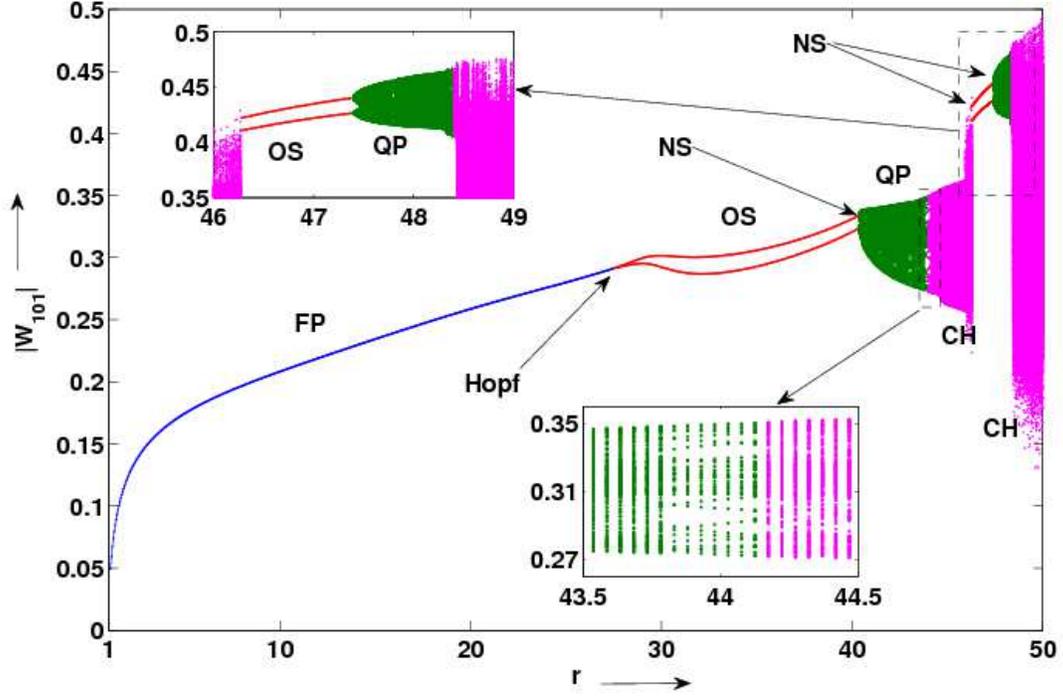}
\end{center}
\caption{Bifurcation diagram computed using the 30-mode model of Paul {\em et al.}~\cite{Paul:highP}. ÔFPÕ (blue curve) is the steady convection roll, `OS' (red curve) is the time-periodic roll, `QP' (green patch) is the quasiperiodic roll, and `CH' (pink patch) is the chaotic state. `NS' indicates the Neimark-Sacker bifurcation point. A window of periodic and quasiperiodic states is observed in the band of $r = 46.2:48.4$.  This bifurcation diagram has certain similarities with that computed using DNS results (Fig.~\ref{fig:bif_W}).}
\label{fig:bif_model_W}
\end{figure}

\begin{figure}[t]
\begin{center}
\includegraphics[height=!,width=16cm]{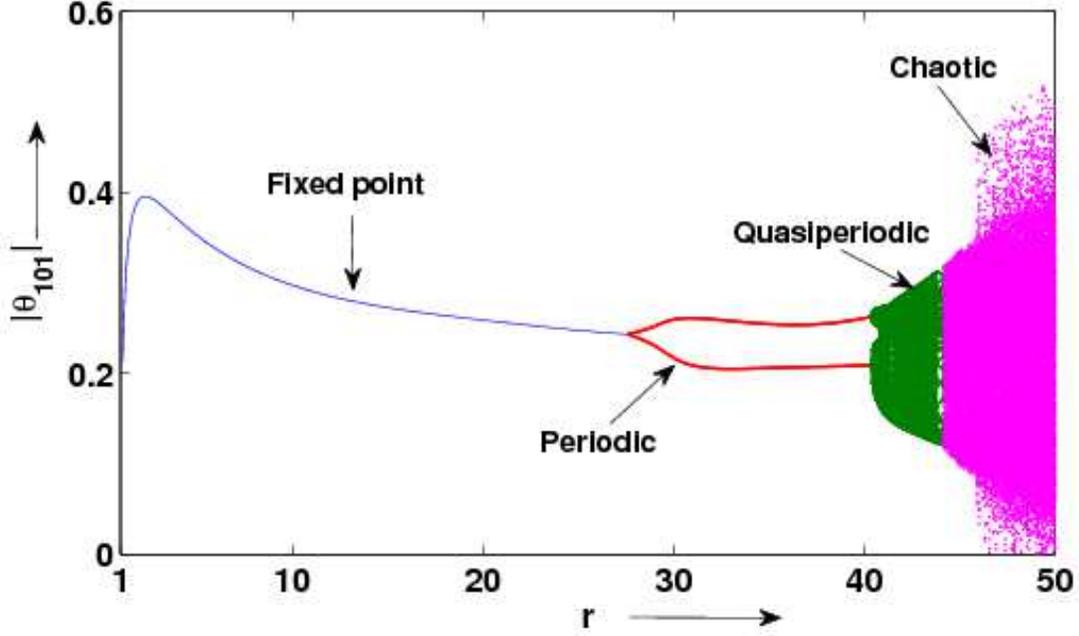}
\end{center}
\caption{Bifurcation diagram for $|\theta_{101}|$ computed using the 30-mode model of Paul {\em et al.}~\cite{Paul:highP}. Various bifurcations are similar to as in FIG.~\ref{fig:bif_model_W}. However this bifurcation diagram is for a different initial condition, which does not show a window of periodic and quasiperiodic states after chaos. This illustrates the coexistence of periodic and chaotic states in the range $r = 46.2:47.4$.  For comparison with DNS bifurcation diagram, see Fig.~\ref{fig:bif_theta}).}
\label{fig:bif_model_theta}
\end{figure}

In the present paper we focus on explaining the origin of various convective patterns using detailed bifurcation analysis.  Recently Paul~{\it et al.}~\cite{Paul:highP} performed similar studies using a 30-mode model  that was derived using  a Galerkin projection of 2D RBC on appropriate large-scale modes.  The features presented here are quite similar to the bifurcation scenario presented for the model (Figs.~\ref{fig:bif_model_W} and \ref{fig:bif_model_theta}).  There are however some differences, e.g., the low-dimensional model has limit cycles and a chaotic attractor as coexisting attractors, while the DNS shows fixed points and a chaotic attractor as coexisting attractors. Another difference stems from the fact that for a given pattern, the $r$ values for 2D DNS is larger than the corresponding model values.  

Experiments, DNS, and low-dimensional models reveal strong dependence of the convective patterns and emergence of chaos on the Prandtl number and the aspect ratio.  Our DNS results are for $P=6.8$ and aspect ratio of $2\sqrt{2}$.   We need to perform more extensive DNS to come up with a comprehensive picture of large-P convection. After the discussions on bifurcation diagrams, we describe the dynamics of chaotic states in the next section.

\section{Dynamics of the chaotic states}\label{sec:chaos}

\begin{figure}[t]
\begin{center}
\includegraphics[height=!,width=16cm]{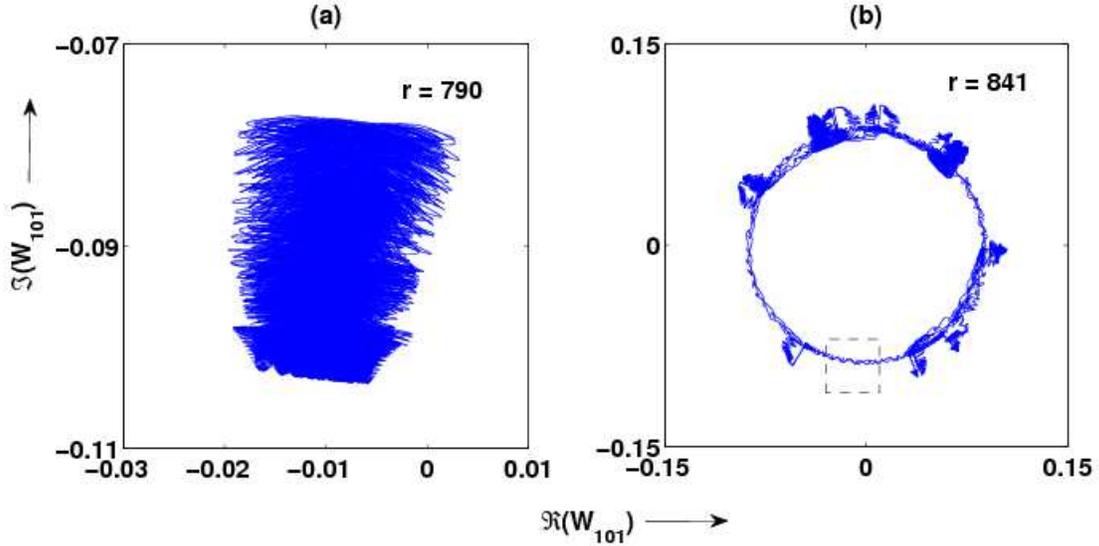}
\end{center}
\caption{Phase space projection of the chaotic attractor on the $\Re(W_{101})-\Im(W_{101})$ plane for (a) $r=790$ for which the phase of the mode $W_{101}$ lies in the band of $(1.43\pi:1.51\pi)$; (b) $r=841$ for which the phase of $W_{101}$ takes all values in $(0,2\pi)$.  The range of the chaotic attractor of figure (a) is depicted by the dashed box of figure (b). }
\label{fig:chaos_phasespace}
\end{figure}

As described in the previous section, chaos appears in 2D RBC for $P=6.8$ and $\Gamma=2\sqrt{2}$ through a quasiperiodic route.  The chaotic attractor for $750 \le r \le 840$ is confined in a narrow region of the phase space, while the region of the chaotic attractor  for $841 \le r \le 849$ is much larger.   This difference between the sizes of the chaotic attractors can be understood by projecting the phase space on $\Re(W_{101})- \Im(W_{101})$ plane (see Fig.~\ref{fig:chaos_phasespace}).  The figure in the left panel corresponds to the narrower chaotic attractor for $r=790$, similar to Fig.~\ref{fig:phasespace}(e).  In the left panel, the phase space trajectories for $r=790$ wander about chaotically in the region $\Re(W_{101}) =(-0.019:0.0024) , \Im(W_{101}) = (-0.10:-0.077)$, with the phase of the Fourier mode $W_{101}$ confined in a narrow band of $(1.43\pi:1.51\pi)$ radians.  Note that the phase of the Fourier mode $W_{101}$ could take any mean value due to azimuthal symmetry discussed in the previous section.     
 
 For $841 \le r \le 849$,  the chaotic attractor increases in size as a result of an ``attractor-merging crisis''.  A sample of such a chaotic attractor is shown in the right panel of Fig.~\ref{fig:chaos_phasespace} for $r=841$.  The system evolves in such a way that the phase of the mode $W_{101}$ varies in a  narrow band for a while, and then the system  jumps abruptly to another band of phases.   Here, the phase of the mode $W_{101}$ can take any value between $0$ and $2\pi$ unlike the smaller chaotic attractor described above.  The region occupied by the phase space attractor at $r=790$ is confined within the dashed box of the right panel of Fig.~\ref{fig:chaos_phasespace}.
 
\begin{figure}[t]
\begin{center}
\includegraphics[height=!,width=15cm]{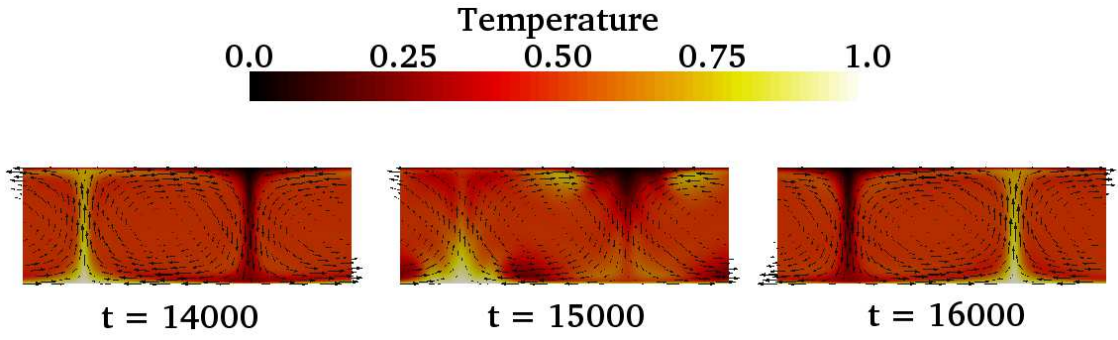}
\end{center}
\caption{The snapshots of the chaotic travelling rolls at $r=841$ and $P = 6.8$. The convective rolls move to the left. }
\label{fig:trav_rolls}
\end{figure}

 The above mentioned abrupt changes in the phase of the Fourier mode $W_{101}$ correspond to sudden or chaotic movements of the convective rolls along the horizontal direction~\cite{Paul:pramana_2010}.  We show this feature in Fig.~\ref{fig:trav_rolls}, where we have drawn the snapshots of the flow patterns for $r=841$. The three frames of Fig.~\ref{fig:trav_rolls} represent $t=14000$, just before the rolls start moving,  $t=15000$, during the movement of the roll, and  $t=16000$, when the roll has moved.  As mentioned earlier, the movements of the rolls are quite abrupt.    A movie depicting the movement of the roll during this time interval  is available at~\cite{movie}.
 
\begin{figure}[t]
\begin{center}
\includegraphics[height=!,width=16cm]{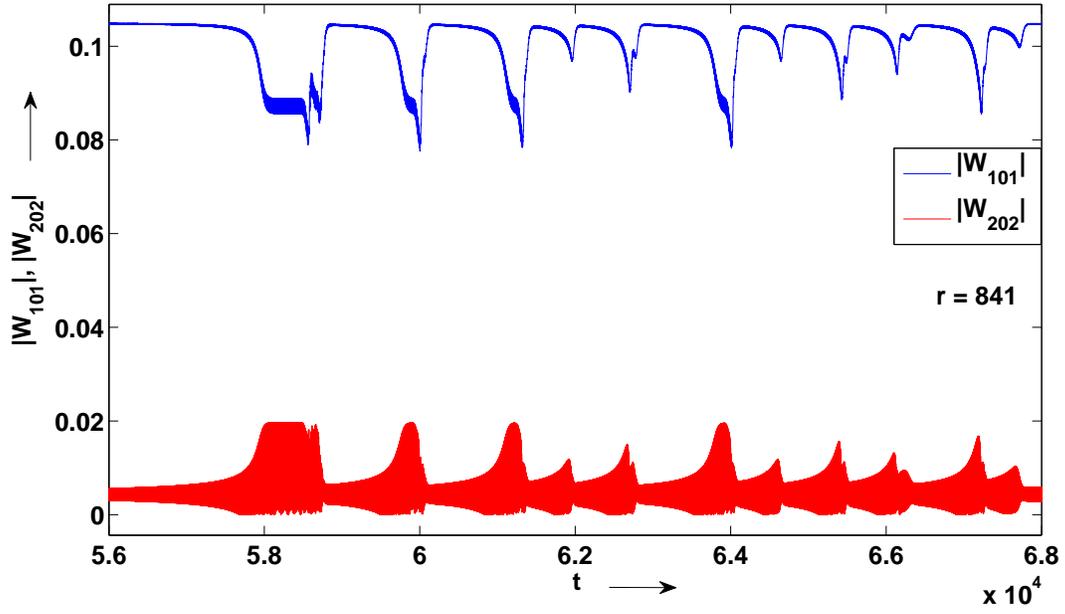}
\end{center}
\caption{Time series of $|W_{101}|$ and $|W_{202}|$ modes for $r=841$.  $|W_{101}|$ value decreases sharply, while $|W_{202}|$ rises when the phase of $W_{101}$ changes abruptly. }
\label{fig:w101_w202_reversal}
\end{figure}

 The travelling rolls are also related to the reorientations of convective structures and wind reversals observed in experiments and simulations~\cite{Cioni:JFM, Brown:JFM, Mishra:JFM}.   We also observe interesting patterns in the energy content of the dominant Fourier modes during these abrupt events.  The time series of $|W_{101}|$  shown in Fig.~\ref{fig:w101_w202_reversal}  shows a decrease in the values of $|W_{101}|$ up to around 20\% of its maximum value when the phase of the Fourier mode changes abruptly.   The value of $|W_{202}|$ however increases suddenly during these events.  These features of the energetics of the Fourier modes are similar to those observed in the DNS of wind reversals in a cylindrical container \cite{Mishra:JFM}.
 
As described earlier, convective states like steady, periodic, quasiperiodic, and chaotic rolls can be conveniently described using the large scale or small wavenumber Fourier modes.  Variation of the magnitudes of these Fourier modes with $r$ follows some interesting patterns  that will be described in the next section.

\section{Properties of the large-scale modes}\label{sec:largescale}

\begin{table}
\begin{tabular}{|c|ccc|ccc|}
\hline
 & & $|W_{101}|$ & &  & $|\theta_{101}|$ & \\
\hline
$r$~  & ~DNS & model & Lorenz~  & ~DNS & model &  Lorenz~\\
\hline
1.2	&  0.057	& 0.068 & 0.066	&  0.28	& 0.28 & 0.27 \\
1.7	&  0.10	& 0.11 & 0.10	&  0.38	& 0.38 & 0.36\\
4	&  0.17	& 0.16 & 0.14	&  0.37	& 0.37 & 0.32\\
6	&  0.19	& 0.18 & 0.15	&  0.33	& 0.34 & 0.27\\
8	&  0.20	& 0.20 & 0.15	&  0.30	& 0.31 & 0.24\\
10	&  0.20	& 0.21 & 0.15	&  0.28	& 0.30 & 0.22\\
20	&  0.22	& 0.26 & 0.16	&  0.22	& 0.26 & 0.16\\
\hline
\end{tabular}
\caption{Comparison of values of $|W_{101}|$ and $|\theta_{101}|$ from the DNS, the 30-mode model and the Lorenz model. The values match closely for low $r$. With higher $r$ the values start to differ significantly.}\label{table:DNS_model}
\end{table}

\begin{figure}[t]
\begin{center}
\includegraphics[height=!,width=16cm]{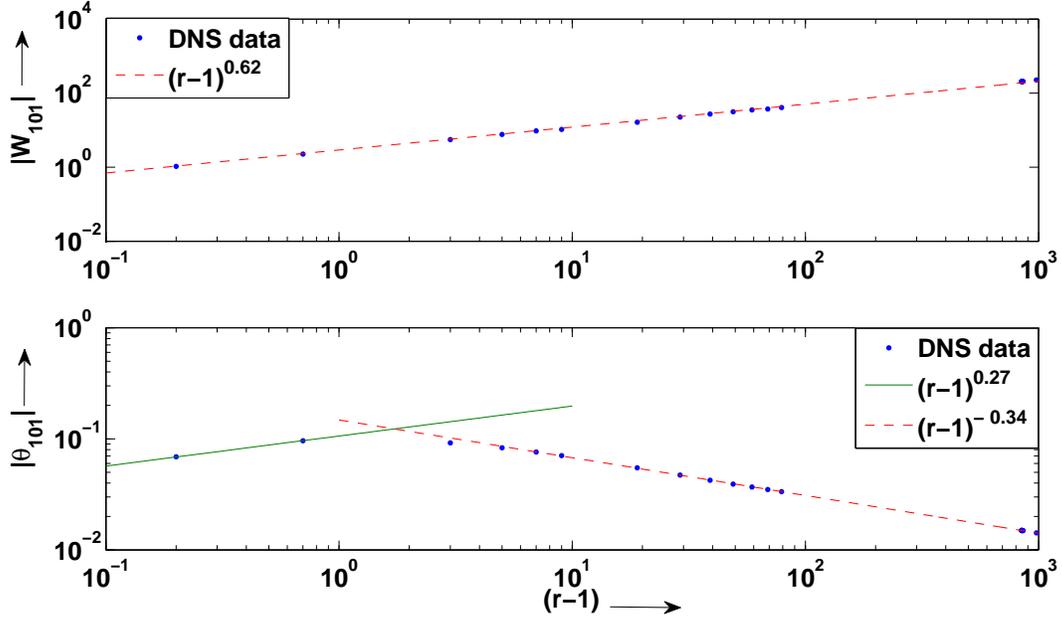}
\end{center}
\caption{Variation  of $|W_{101}|$ and $|\theta_{101}|$ as a function of $(r-1)$ obtained from the DNS results. $|W_{101}| \sim (r-1)^{0.62}$, while $|\theta_{101}| \sim (r-1)^{0.27} $ for $r < 2$, and $|\theta_{101}| \sim (r-1)^{-0.34} $ for $ 2 < r < 1000$.}
\label{fig:w11_t11_DNS}
\end{figure}

In Fig.~\ref{fig:w11_t11_DNS} we plot the values $|W_{101}|$ and $|\theta_{101}|$ for the stable fixed points as a function of $(r-1)$.  We observe that  $|W_{101}| \sim (r-1)^{0.62}$ for $r$ up to 1000.  However $|\theta_{101}|$ shows two distinct scaling: $\sim (r-1)^{0.27}$ for  $1 \le r < 2$,  and $\sim (r-1)^{-0.34}$ for $2 < r < 1000$.   We also compare the values of $|W_{101}|$ and $|\theta_{101}|$ computed using DNS, the 30-mode model of Paul~{\em et al.} \cite{Paul:highP}, and the Lorenz model~\cite{hilborn:book}  for $P=6.8$.  The entries listed in Table~\ref{table:DNS_model} indicate that the results of the 30-mode model matches quite well with DNS for $r \le 10$, while Lorenz model works well for $r$ only up to 4.  For $r$  far away from the onset, many Fourier modes get excited, and the low-dimensional models are not expected to match with DNS quantitatively.  However the energetic Fourier modes provide fairly good  qualitative description as visible from the significant similarities between the DNS bifurcation diagrams (Figs.~\ref{fig:bif_theta} and \ref{fig:bif_W}) and the corresponding diagrams obtained from the low-dimensional model (Figs.~\ref{fig:bif_model_W} and \ref{fig:bif_model_theta}).

\begin{figure}[t]
\begin{center}
\includegraphics[height=!,width=16cm]{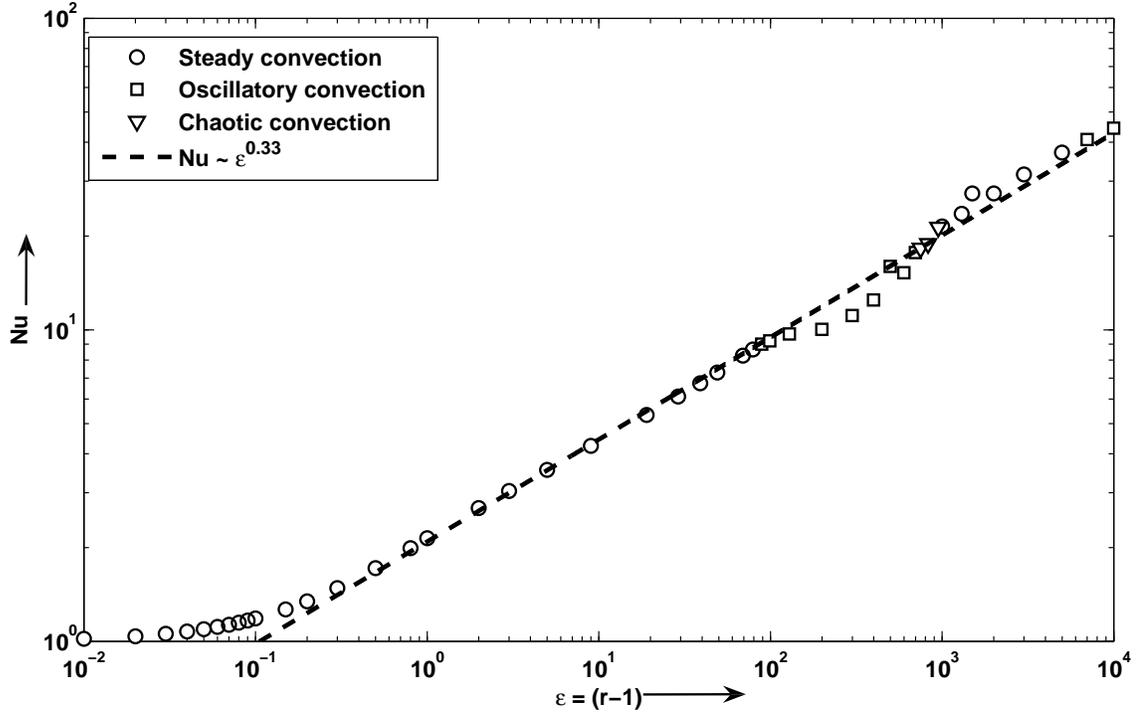}
\end{center}
\caption{Variation of Nusselt number $Nu$ as a function of $(r-1)$.   $Nu \sim (r-1)^{0.33 \pm 0.01}$ for $r>2$.}
\label{fig:Nu}
\end{figure}

Another quantity of interest related to the large-scale convection is the Nusselt number, which is the ratio of the total heat flux and the conductive heat flux.  Experiments, numerical simulations, and phenomenological theories project a power law behaviour of the Nusselt number as a function of the Rayleigh number \cite{ahlers:RMP_2009}.  The predicted spectral indices for various conditions are 1/4, 2/7, 3/10, 1/3, 1/2, etc.~\cite{ahlers:RMP_2009}.  The Nusselt numbers $Nu$ for our various DNS runs are shown in Fig.~\ref{fig:Nu} where we plot  $Nu$ as a function of $(r-1)$. For $r >2$,  $Nu\sim (r-1)^{0.33\pm 0.01}$ with an observable deviation from the fit for $100 \lesssim r \lesssim 600$.  These results are in good agreement with the experimental findings of Cioni~\cite{Cioni:JFM} and Niemela {\it et al.}~\cite{Niemela_2000}, as well as many numerical simulations, e.g., by Moore and Weiss~\cite{moore_weiss:jfm_1973}.  This agreement also acts as a validation of our numerical code.


\section{Conclusions}
In this paper we perform direct numerical simulations of two-dimensional RBC for $P=6.8$, aspect ratio $\Gamma=2\sqrt{2}$, and the reduced Rayleigh number $r$ up to $5\times 10^5$ ($R = 3.3 \times 10^7$).    Simulations reveal various convective states like steady, periodic, quasiperiodic, and chaotic rolls.  Using simulation results we construct bifurcation diagrams that help us understand the origin of the observed convective patterns.  The steady convective rolls are born through a pitchfork bifurcation at $r=1$.  At around $r \simeq 80$, time-periodic rolls appear through a Hopf bifurcation. The periodic state bifurcates to period-2 state through a period-doubling bifurcation. Through a Niemark-Sacker bifurcation, these period-2 rolls turn into quasiperiodic rolls which in turn becomes chaotic through a quasiperiodic route to chaos.  The chaotic attractors undergo an ``attractor-merging crisis'' to generate a larger chaotic attractor  involving abrupt changes in the phase of the Fourier modes leading to abrupt motion of the convective rolls.  In our DNS we also observe coexistence of stable fixed points and a chaotic attractor as a result of an inverse subcritical Hopf bifurcation.  After a while, only stable fixed points survive due to the disappearance of the chaotic attractor through a ``boundary crisis''.  These fixed points subsequently bifurcate to periodic and chaotic states through successive bifurcations. 

Several of the above mentioned convective structures have been observed earlier  in some experiments~\cite{gollub:jfm_1980,Maurer:1979,Libchaber}, simulations~\cite{curry:jfm_1984,mclaughlin_orszag:jfm_1982,Mukutmoni:JHT_1993,NishiYahata:DNS,Yahata:DNS}, and low-dimensional models~\cite{Yahata:lowD_3freq,Paul:highP}.  However, we observed for the first time stable fixed points beyond chaos, coexistent stable fixed points and chaotic state, and  several occurrences of ``crisis''.  The above mentioned bifurcation diagrams have similarities with those derived for Paul~{\it et al.}'s 30-mode model for the same Prandtl number \cite{Paul:highP}. 

Experiments of Gollub and Benson \cite{gollub:jfm_1980} exhibit other kinds of patterns and chaos for different sets of Prandtl numbers (2.5, 5) and aspect ratios (2.4, 3.5).  We  are exploring these ranges of Prandtl numbers and aspect ratios through both DNS and low-dimensional models and hope to achieve a more comprehensive picture of large-P convection in future.
The results in our paper illustrate the usefulness of bifurcation diagrams in understanding the origin of various convective patterns.  

\begin{acknowledgments}
We thank K. R. Sreenivasan, J.  Niemela, Daniele Carati, Arul Lakshminarayan, and Pinaki Pal for discussions and  important suggestions.  We thank Computational Research Laboratory for providing computational resources to complete this work. Part of the work was supported by the grant of Swarnajayanti fellowship by Department of Science and Technology, India.  
\end{acknowledgments}


\pagebreak


\end{document}